\documentclass{webofc}
\usepackage[varg]{txfonts}   

\usepackage{footmisc}

\usepackage{lineno}

\def\rooteve{ROOT-Eve\xspace}
\def\reve{REve\xspace}
\def\rcore{RenderCore\xspace}
\def\three{{\small THREE.js}\xspace}

\def\stt#1{{\small\texttt{#1}}}
\def\etal{\emph{et al.}\xspace}

\def\postfigskip{\vskip-4mm}

\begin{document}

\title{\rcore – a new WebGPU-based rendering engine for ROOT-EVE}

\author{
       \firstname{Ciril} \lastname{Bohak}\inst{4,5}
  \and \firstname{Dmytro} \lastname{Kovalskyi}\inst{2}
  \and \firstname{Sergey} \lastname{Linev}\inst{3}
  \and \firstname{Alja} \lastname{Mrak Tadel}\inst{1}
  \and \firstname{Sebastien} \lastname{Strban}\inst{4}
  \and \firstname{\mbox{Matev\v{z}}} \lastname{Tadel}\inst{1}\fnsep\thanks{\email{mtadel@ucsd.edu}}
  \and \firstname{Avi} \lastname{Yagil}\inst{1}
}

\institute{UC San Diego, La Jolla, CA, USA 92093
  \and     MIT, Cambridge, MA, USA, 02139
  \and     GSI, Darmstadt, Germany 64291
  \and     University of Ljubljana, Ljubljana, Slovenia 1000
  \and     KAUST, Thuwal, Saudi Arabia, 23955
}

\abstract{
\rooteve (\reve), the new generation of the ROOT event-display module, uses a web server-client model to guarantee exact data translation from the experiments' data analysis frameworks to users' browsers. Data is then displayed in various views, including high-precision 2D and 3D graphics views, currently driven by \three rendering engine based on WebGL technology.

\rcore, a computer graphics research-oriented rendering engine, has been integrated into \reve to optimize rendering performance and enable the use of state-of-the-art techniques for object highlighting and object selection. It also allowed for the implementation of optimized instanced rendering through the usage of custom shaders and rendering pipeline modifications. To further the impact of this investment and ensure the long-term viability of \reve, \rcore is being refactored on top of WebGPU, the next-generation GPU interface for browsers that supports compute shaders, storage textures and introduces significant improvements in GPU utilization. This has led to optimization of interchange data formats, decreased server-client traffic, and improved offloading of data visualization algorithms to the GPU. FireworksWeb, a physics analysis-oriented event display of the CMS experiment, is used to demonstrate the results, focusing on high-granularity calorimeters and targeting high data-volume events of heavy-ion collisions and High-Luminosity LHC. The next steps and directions are also discussed.
}

\maketitle

\section{Introduction}
\label{intro}

The new event visualization environment of the ROOT data processing framework \cite{root-paper} ~---~ called \rooteve or just \reve ~---~ is a rewrite of the now more than 15-years old components, TEve \cite{root-teve} and TGL \cite{root-tgl}, for the modern times using web client--server application model, web-based graphical user interfaces and 2D \& 3D graphics. The code and APIs have been modernized in line with the ROOT-7 campaign. On the client side, \reve is based on OpenUI5 JavaScript application framework \cite{openui5-sw} and JSRoot \cite{jsroot,jsroot-sw}.

Development of \reve has been driven by the needs and requirements of the Fire\-works\-Web application, itself a rewrite of Fireworks, the physics-analysis oriented event-display of the CMS experiment \cite{fworks-1,fworks-2}. Since ROOT version 6 provides a fully \stt{C++} standard-compliant interpreter, \stt{rootcling}, including support for lambda expressions, it was possible to port several high-level Fireworks features into \reve: support for physics collections \& physics items, physics item filtering, and table-views with custom column expressions. Proof-of-concept implementation and design choices for \reve and FireworksWeb were presented at CHEP-2018 \cite{fworksweb-chep18} (client--server core, data exchange, remote method execution) followed a year later, at CHEP-2019, by a functional prototype of CMS FireworksWeb \cite{fworksweb-chep19}. Towards the end of 2021, \emph{event-display service} for CMS was deployed at CERN and at UC San Diego. Access to any CMS data is available through AAA, the CMS XRootd data federation, from CERN EOS, and from CERNBox. To improve the data access locality at UCSD, the AAA access is fronted by an XCache instance. \reve is the core technology used for CMS visualization. The legacy TEve-based application is still supported for online event visualization in the control room at P5, geometry browser for simulation geometry development and debugging, and to run custom user extensions that have not been ported yet to the new framework.

JSRoot uses \three \cite{three.js} to display 3D plots and detector geometry and was initially used by \reve without much afterthought --- it was seen as a very general and well-supported option. The fact that \reve could use an existing rendering library was even seen as one of the benefits of adopting the web as the user interaction layer. However, as development reached the level of adding advanced features that were supported in the custom TEve / TGL low-level OpenGL-1.x rendering engine \& scene graph library, several complications arose, as will be discussed in section \ref{sec:three-vs-rcore}.

Around that point, a predecessor Med3D~\cite{med3D} to \rcore, a lightweight JavaScript WebGL-2.0 rendering engine, was presented at the HEP Software Foundation workshop in Naples (2018) with an expression of interest for collaboration with the HEP community by the Laboratory for Computer Graphics and Multimedia at Faculty of Computer and Information Science at the University of Ljubljana. While the Med3D was intended for collaborative medical visualization on the web, its flexible design allowed easy adaptation for other use cases. \rcore, is even more adaptable, especially in creating custom rendering pipelines and supporting arbitrary data to be used within them, making it much more adaptable than \three.  The first expression of interest from the \reve side was made in 2019 with some in-depth explorations with partial \reve-side implementation following through 2020 and 2021 when the final commitment to include \rcore in \reve was made from both sides and \rcore software was made publicly available \cite{rcore}. Since the end of 2022, \rcore is the default render engine for \reve.

\section{Motivation for migration to \rcore}
\label{sec:three-vs-rcore}

As mentioned in the introduction, TEve used a custom OpenGL-1.x engine developed in sync with TEve as needed to support the CAD-like features of EVE: overlays, overlay event-handling, overlay GUI elements, high-precision selection, and well-pronounced object outlines and highlights. Controlling the low-level details of the render flow offered great flexibility in implementing these advanced features that are typically unavailable in standard rendering engines. Before \reve, migration to a modern, shader-based OpenGL was never even considered, as it would have required a major rewrite of all components. Client--server architecture of \reve, however, mandates serialization of visualization data with complete decoupling of state between server- and client-side representations. The required rewrite of visualization data structures and their generating algorithms was then also used to try to express these structures so that memory buffers from the server (\stt{C++}) side can be sent directly to memory buffers in JavaScript, and then passed on as render-buffers to WebGL without any reinterpretation or even reshuffling. These two topics, the ability to implement advanced features and efficient data passing, are at the core of the issue of selecting a rendering engine, using it, and extending it for implementation of an even-display framework. In this section, we discuss the reasons for the transition from a large, fully-featured, community-driven rendering engine to a lightweight, research \& teaching-driven one.

\paragraph{High-level issues with \three}

\three is a large project aiming to provide a cross-browser, general-purpose 3D library. As such it has a lot of functionality and features \reve does not need. It has an extensive user base, and as it also aims to be an easy-to-use framework, the features and functionality need to be tightly integrated from API classes through the rendering pipeline all the way to shader implementation codes. This means that any significant changes need to be made in several places and coexist with several other supported features, however exotic they might be from one's perspective. As we have been learning modern web-based graphics during the early development of \reve, introducing changes into \three or even just extending it with custom algorithms and functionality turned out to require great diligence as implementation often required an understanding of several components and layers of the library as well as the corresponding computer graphics principles.

The same argument works against us also once changes are successfully implemented: there is no desire from \three maintainers to include them in the main repository as the CAD-like features that \reve requires are not used by other library users. Therefore, including such changes is disruptive because it complicates the code, as explained above, and introduces potential maintenance, support, and documentation issues. Even our simpler changes, related to the usage of geometry buffers that would allow us to avoid copying data in JavaScript code, were rejected as not being of interest to a wider community. This resulted in our changes being kept in the \reve repository, some as run-time patches for \three code.

At the same time, the release and low-level change rate for \three is rather large, and while API classes are guaranteed to remain supported to some extent, the back-end code can and does change significantly between releases. For \reve, complete stability of rendering code is preferred as very limited developer time is available for its maintenance. For example, ROOT's TGL package used by the legacy TEve has been stable since 2008.

\paragraph{Unresolved technical \& functional issues with \three}

As mentioned above, it was challenging to implement advanced features that spawn across multiple levels of the \three framework. An example of such a feature is support for memory-optimized, morphable instantiated objects, e.g., polygons or polyhedra with some varying properties such as position, angle(s), scale(s), and color, that are required for visualization of high-volume simulation and reconstruction data. Implementing this feature would require support in API classes, render-driver pipeline, and shaders, including custom shader input on a per-primitive level. 

Another problem was picking or selecting an object or its component drawn at a given window position. \three uses ray–mesh intersection for this purpose. This does not work well for point data (sprites) and lines, especially at high levels of magnification required for visualization of vertex regions. Further, \reve uses object and sub-object outlines (highlights) to signal selection contents and under-the-mouse objects associated with the pop-up object details window. We had a workaround solution for object outlining but never adequately resolved the picking problem.

3D lines of arbitrary thickness are challenging in any post-OpenGL-1.x framework as thick lines are no longer required to be supported by the standard. Things are even more complicated by the need to perform pixel-correct line picking and draw the outlines. This has never been addressed in the \three implementation.

\medskip

In hindsight, it is clear that \reve requirements for a rendering engine, coupled with our inexperience with modern graphics frameworks, led us to underestimate the required time to implement the required functionality and overestimate the ease of implementation of the mentioned features. It also took us a while to realize and to come to terms with the fact that using \three, an off-the-shelf community-based project, does not work in our case.

\paragraph{Adoption of \rcore}

The two previous sub-sections list our issues with the \three-based implementation of \reve. All of them have been successfully addressed in the \rcore implementation. This does not mean the implementation was easy, but it was certainly easier and, in fact, possible. Some of the implementation details are discussed in the next section.

Collaboration on \reve--\rcore integration was extremely productive and efficient. It was of great help for \reve developers to be able to discuss details of \rcore code, to plan together the required extensions, and to be actively involved in the implementation of low-level changes in \rcore. This work taught them the art of the modern shader-based rendering frameworks, provided meaningful extensions to \rcore, and implemented advanced features in the \reve--\rcore interface.


\section{REve@\rcore: implementation highlights}

\paragraph{Object \& sub-object picking}

Object \& sub-object picking got implemented via rendering to off-screen buffers (single-channel 32-bit unsigned integer). This method gives pixel-perfect results but has to be implemented via dedicated shaders (or code paths in standard shaders) that put the object's unique identifier into the output buffers. The identifiers are assigned automatically during the pre-render traversal. A limited viewport, 32x32 pixels around the pick position, reduces memory usage and rendering time. View-space z-coordinate is also extracted (as a 32-bit floating-point number) to place annotation anchors and set the camera rotation center.

Sub-object or instance picking is implemented as a two-stage process. After the primary object is selected, that object alone is drawn again with a special shader flag that makes it use either instance-identifier (for instanced draw) or vertex-identifier (for mesh-encoded instancing) as the output identifier. Optimized implementation required changes in the \rcore mesh renderer code and a custom implementation of two-pass selection in \reve render-driver as an interpretation of picking results depends on implementation details of \reve client-side objects.

\paragraph{Object outlines}

Outline rendering is based on a custom implementation of rendering into textures to extract a subset of geometry buffers traditionally used for deferred shading. View-space depth, view-space normal vectors, and view-direction vectors are used to determine the selected object's edges and sudden changes in the normal direction. This allows for the outlining of the object's silhouette as well as the marking of the inner edges to aid in shape recognition.

\begin{figure}[thb]
\centering
\includegraphics[width=0.6\hsize]{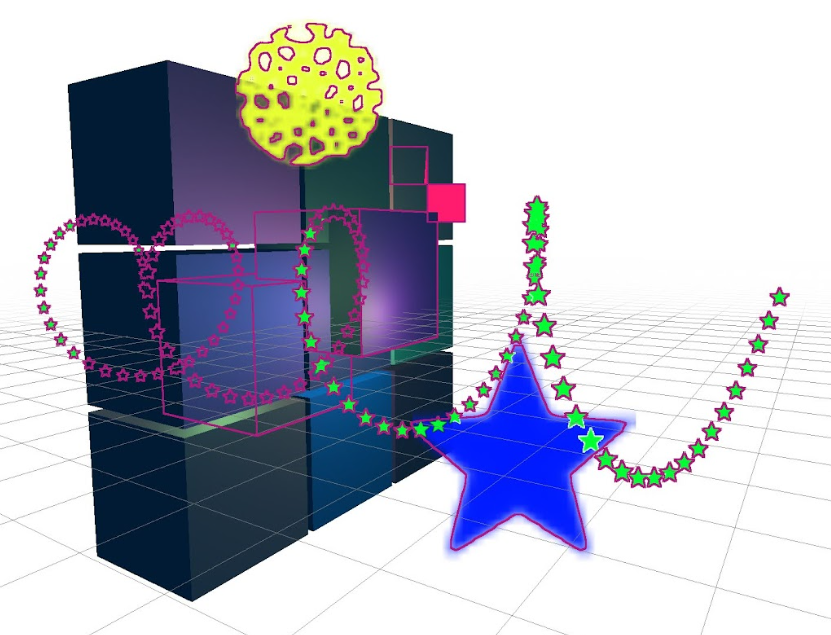}
\postfigskip
\caption{\rcore test for outline drawing. The selected cubes also have edge outlines. Sprites (used for markers in \reve) are outlined in accordance with transparency of the defining texture --- a star marker is outlined as a star despite being drawn as a textured rectangle.}
\label{fig:outlines}
\end{figure}

\begin{figure}[thb]
\centering
\includegraphics[width=0.9\hsize]{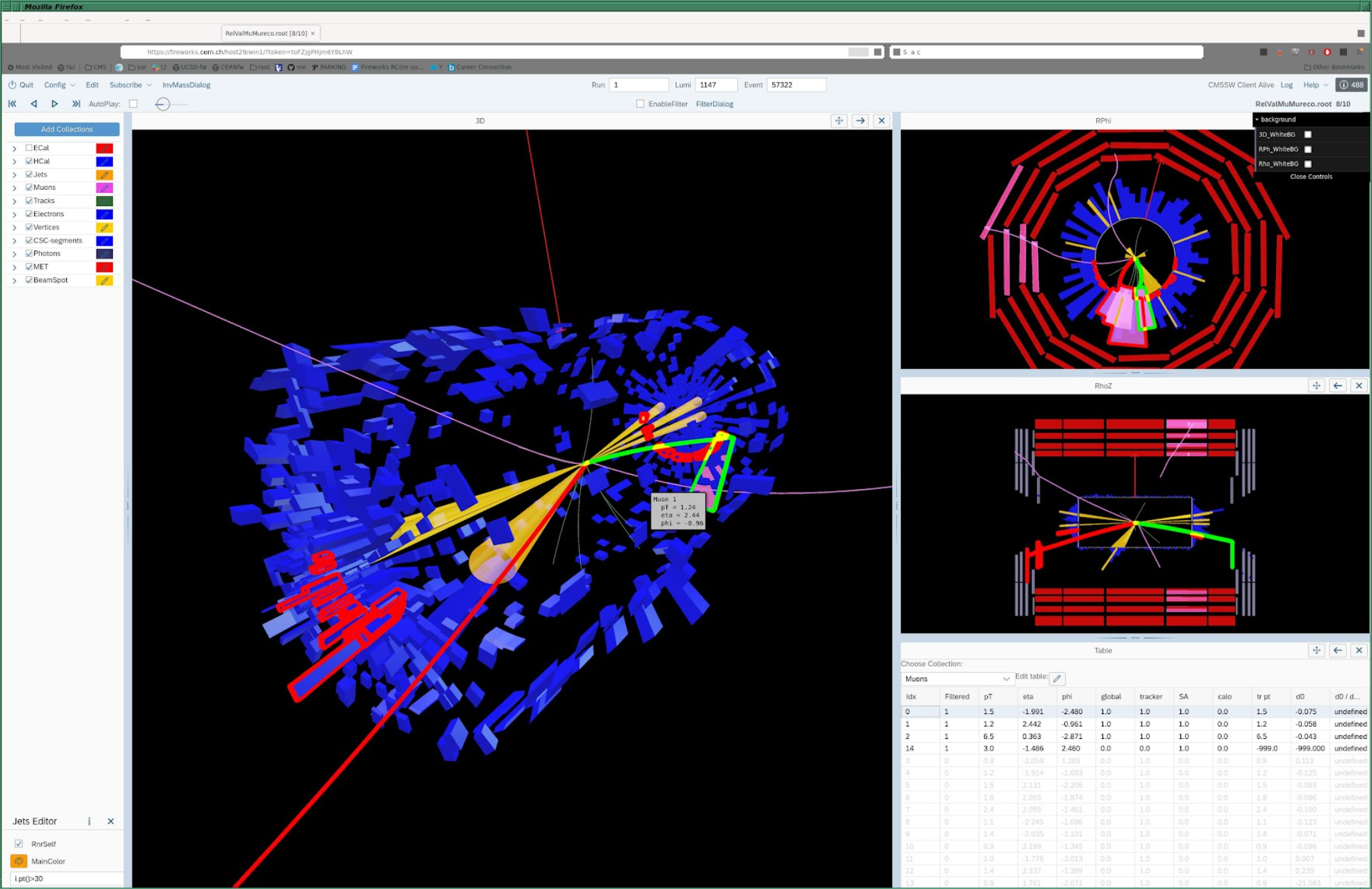}
\postfigskip
\caption{Selection and highlight in \reve. Two selected objects are outlined in red: a muon and a sub-selection of calorimeter towers. A highlighted (under the mouse) muon is outlined in green. Notice also the pop-up information box showing track parameters.}
\label{fig:sel-hilite}
\end{figure}

\reve supports multiple lists of selected objects (or sub-objects), each having its own outline colors and outline widths. Each selection list must be processed separately, and all the outlines must be accumulated into the final output buffer. Tight management and reuse of render buffers are required for a memory-efficient implementation. Figure \ref{fig:outlines} shows an image of outlines as implemented in a \rcore test program, and figure \ref{fig:sel-hilite} shows a composite selection \& highlight rendering in \reve.

\paragraph{Instanced objects}

Instanced objects are used when a large number of physics items need to be drawn, e.g., for the display of digits, hits, or calorimeter towers. Grouping a set of such physics items together into a single object (both in \reve and in \rcore) allows for a significant reduction in object representation, data transport, and data processing overheads. The shared part of the state gets stored in the containing object, and the item-specific information gets encoded in object-specific data buffers, as needed for the specific case. This can be any combination of position, rotation with respect to a fixed axis, scaling, color, or some other parameters affecting the shape of the object.\footnote{Standard computer graphics instancing uses the same mesh (unmodifiable) and a full 4x4 transformation matrix for each instance.}

Instancing as implemented in \reve--\rcore requires the instance data to be packed into "data" textures transferred from the server to the GPU without repacking. However, one has to coordinate three things: a) data packing on the server side; b) texture format \& size interpretation when passing it to WebGL; and c) instance stride and data-texture component interpretation in the shader code. This is rather cumbersome and was one of the motivating factors for considering WebGPU-based \rcore implementation for \reve.

\begin{figure}[thb]
\centering
\includegraphics[width=0.7\hsize]{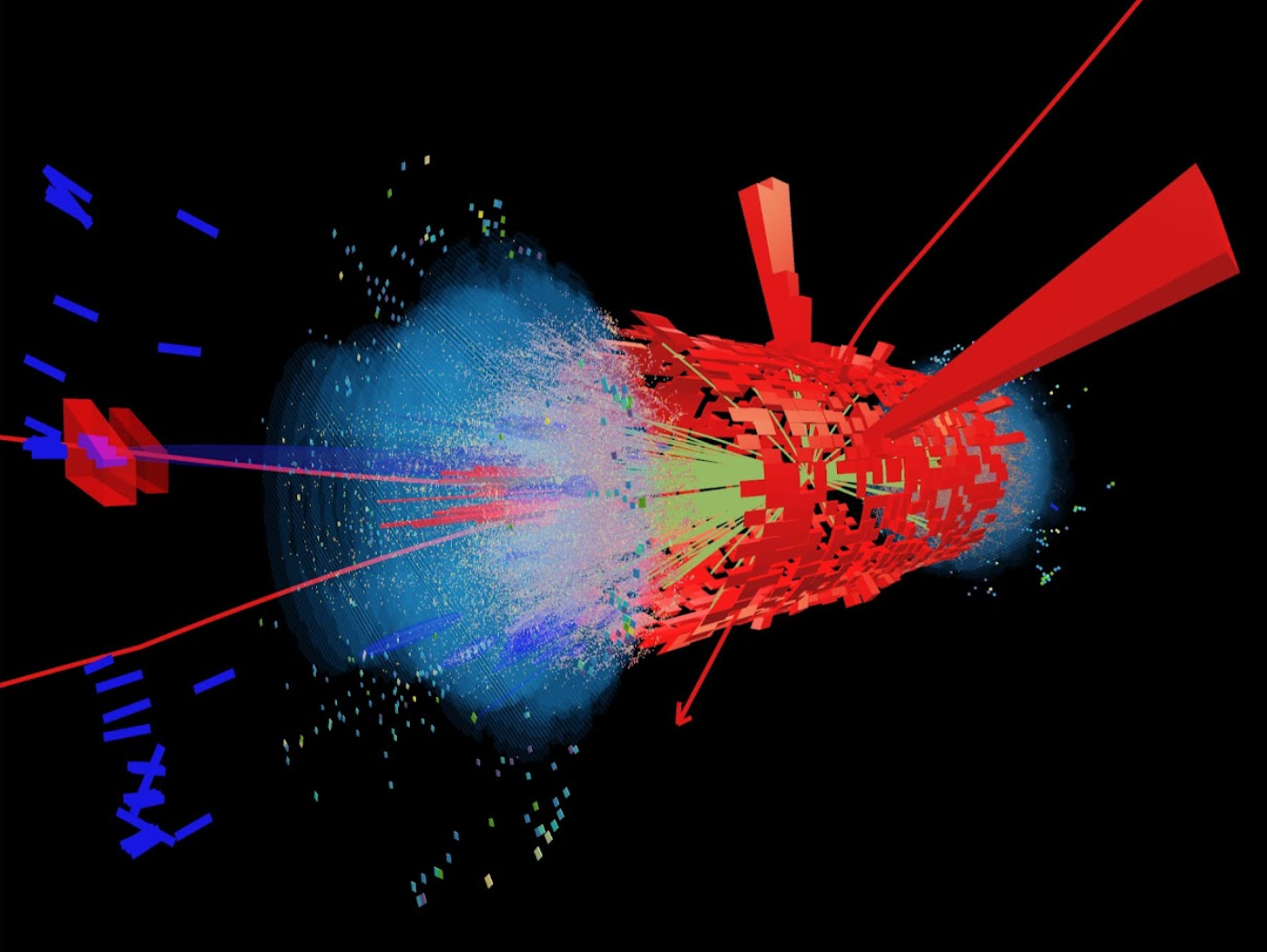}
\postfigskip
\caption{Full view of CMS HGCal visualization. The light-blue halo consists of 1.2 million hexagonal prisms, each rendered as 24 triangles. Interactive refresh rates are achieved on 10-year-old hardware (NVIDIA GTX-970).
}
\label{fig:hgcal-splash}
\end{figure}

\begin{figure}[thb]
\centering
\includegraphics[width=0.7\hsize]{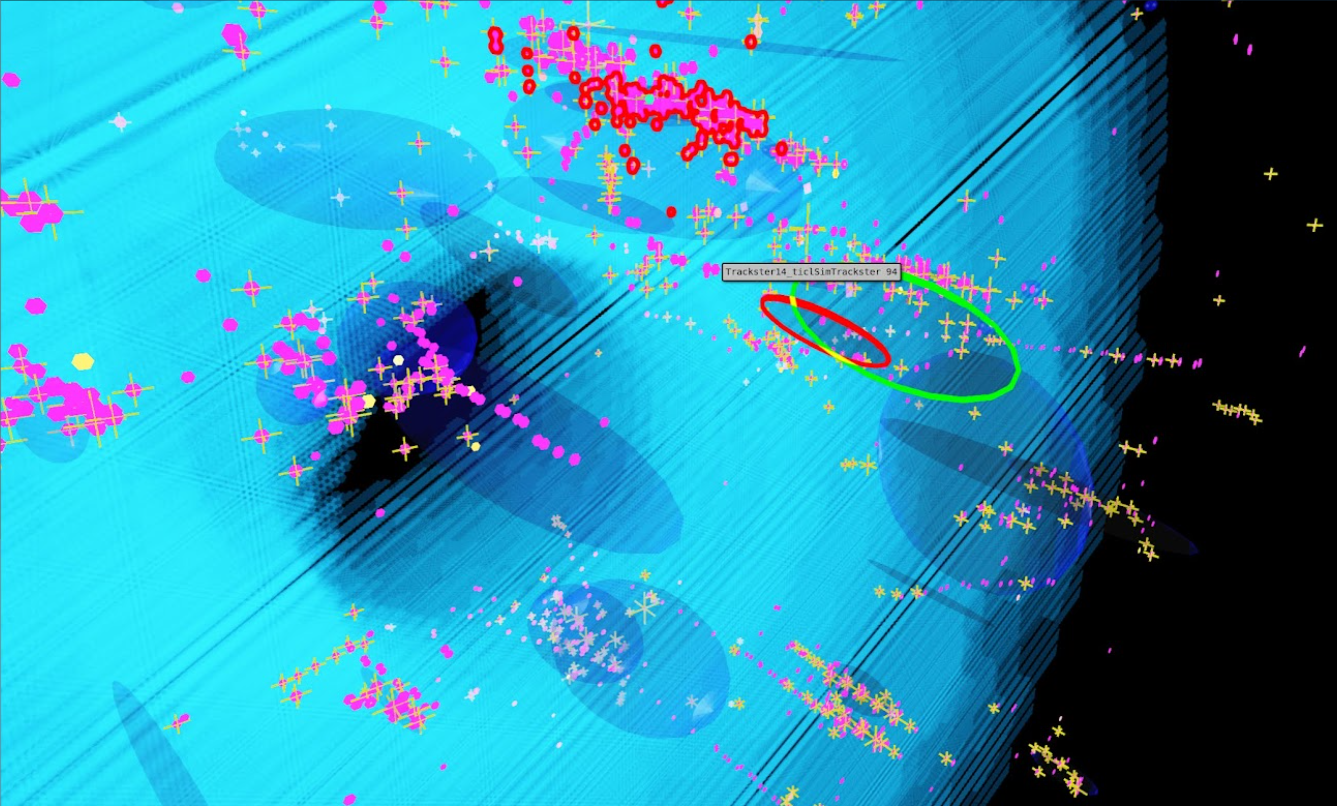}
\postfigskip
\caption{Detail of CMS HGCal visualization showing simulated detector hits and reconstructed clusters and tracklets.}
\label{fig:hgcal-detail}
\end{figure}

Figure \ref{fig:hgcal-splash} shows the visualization of CMS HGCal detector elements implemented via such instancing. Figure \ref{fig:hgcal-detail} shows a detail of the same event where simulated hits and reconstructed physics objects can also be seen.

\section{Ongoing \& future work}

\subsection{Transition to WebGPU}

WebGPU is the upcoming standard for 3D graphics on the web, developed and supported by all major browser vendors. Compared to WebGL, WebGPU introduces significant improvements in GPU utilization as well as support for compute shaders and storage textures, which allow storing to arbitrary positions within shaders without the need for texture sampling. Storage textures, coupled with compute shaders, can potentially be used with structured data to alleviate the need to repack data. Further, WebGPU exposes full control over the rendering pipeline elements and allows for complex state changes to be enacted quickly by switching between several pre-configured pipelines.

\rcore implementation of WebGPU-based rendering is mostly complete. The biggest challenge was the design of an elegant and efficient mechanism for the creation and management of custom pipelines and their coupling with associated materials and shader programs. Appropriate use of render bundles allowed for the recording of render commands and a significant reduction of CPU time usage.

The transition to WebGPU has been motivated by the above performance improvements as well as by the fact that WebGL browser implementations have not received significant updates since 2017 and, in fact, have not been updated beyond the WebGL 2.0 version of the standard, which is built on top of OpenGL ES 3.0\footnote{OpenGL ES is a version of the standard for embedded systems such as mobile devices and is thus additionally limited in comparison to the full version of the standard.}. Additionally, WebGPU utilizes modern back-ends\footnote{Vulkan, Metal, DirectX 12}, which are the same for both embedded and desktop versions. From the \reve side, the main driver was the ability to bind structured data into shaders, therefore eliminating the need to pack data into textures and improving shader-code readability. Further, this could enable us to generate such structures on the fly, along with relevant shader-code fragments, and ship them from the server to the client, allowing for further optimizations and generalizations of instance rendering.

At the time of the conference, the transition of \reve to use WebGPU-based \rcore was on hold because of issues with the GNU/Linux browser support. Since then, most of the issues have been resolved, the transition is resumed, and most core functionalities are already ported.

\subsection{Usage of WebAssembly}

In \reve, complex processing is done at the server during the geometry preparation stage, in \stt{C++}, and final lightweight calculations are performed on the GPU in shaders. This allows us to balance between memory usage, data-transfer volume, and CPU \& GPU usage and bypass some of the bottlenecks typically encountered in web-based visualization. WebAssembly offers additional performance optimization by providing a means to run pre-compiled assembly code within the browser, minimizing the overhead introduced by JavaScript. In addition, WebAssembly allows the development of both desktop and web-based applications in \stt{C++} or \stt{Rust} which can be coupled with WebGPU for rendering, yielding even higher performance in the natively built application while maintaining the same code-base for both web and desktop applications.

We are aiming to explore WebAssembly for some bulk calculations within the scene graph processing (e.g., calculating normal and model-view matrices or performing view-frustum \& clip-plane culling) and for implementation of client-side physics-item filtering.

\section{Conclusion}
\label{sec:conclusion}

RenderCore was selected as the rendering engine for ROOT-Eve. This allowed us to reclaim advanced functionality required for physics-oriented event-display applications and gain low-level control over object-data representation and rendering pipeline. Collaboration with computer graphics professionals made learning low-level details about browser-based 3D graphics and implementing advanced functionality reasonably easy and significantly reduced the required time investment. \reve and \rcore are used as the core visualization technologies for the CMS experiment. The transition of \rcore to WebGPU is ongoing, proceeding in sync with the finalization of the API and implementation in browsers. This will allow us to perform final optimizations of \reve and provide ROOT event visualization with state-of-the-art 3D graphics.

\subsection*{Acknowledgements}

This work was supported by the U.S. National Science Foundation under Cooperative Agreements OAC-1836650 and PHY-2121686.

\end{document}